\documentclass[10pt,a4paper,twoside]{article}

\usepackage{indentfirst}                        
\usepackage{graphicx}                           
\usepackage{amsgen,amsfonts,amssymb,amsbsy}     

\newenvironment{resum}{\begin{quote}\small}{\end{quote}}

\newcommand{\bfsf}[1]{\textsf{\textbf{#1}}}

\def\be{\begin{equation}}
  \def\fe{\end{equation}}
\def\bea{\begin{eqnarray}}
  \def\fea{\end{eqnarray}}

\def\n{\mathrm{n}}         
\def\p{\mathrm{c}}         

\def\a{{\alpha}}   
\def\b{{\beta}}
\def\E{{\mathcal{E}}}

\def\entr{\alpha}
\def\Dv{\Delta}

\def\qaq{{\quad\textrm{and}\quad}}

\def\Lagr{{\mathcal{L}_\mathrm{H}}}
\def\d{\delta}
\def\Gam{\Gamma_{\!\Delta}}
\def\Press{\Psi}

\def\vec#1{\ensuremath{\mathchoice{\mbox{\boldmath$\displaystyle#1$}}
    {\mbox{\boldmath$\textstyle#1$}}
    {\mbox{\boldmath$\scriptstyle#1$}}
    {\mbox{\boldmath$\scriptscriptstyle#1$}}}}

\def\tens{\vec}

\begin{document}

\thispagestyle{plain}           

\begin{center}


  {\LARGE\bfsf{Stationary structure of \\relativistic superfluid neutron stars}}

  \bigskip


  \textbf{R.\ Prix}$^1$, \textbf{J.\ Novak}$^2$ and \textbf{G.L.\ Comer}$^3$


  $^1$\textsl{University of Southampton, UK}\\
  $^2$\textsl{Observatoire de Paris-Meudon, France}\\
  $^3$\textsl{Saint Louis University, USA}.

\end{center}

\medskip


\begin{resum}
  We describe recent progress in the numerical study of the structure of
  rapidly rotating superfluid neutron star models in full 
  general relativity. The superfluid neutron star is described by a
  model of two interpenetrating and interacting fluids, one representing
  the superfluid neutrons and the second consisting of the remaining
  charged particles (protons, electrons, muons).  
  We consider general stationary configurations where the two fluids can  
  have different rotation rates around a common rotation axis. 
  The previously discovered existence of configurations with one fluid
  in a prolate shape is confirmed. 
\end{resum}

\bigskip


\section{Introduction}

Most current studies of neutron star oscillations are based on
a simple perfect fluid model for a neutron star, which neglects
(besides other things like the magnetic field, crust etc.) the
crucial importance of superfluidity. The presence of 
substantial amounts of superfluid matter in neutron stars is backed by
a number of theoretical calculations of the state of matter at these
extreme densities (e.g. see
\cite{baldo92:_superfl_neutr_star_matter,sjoberg76:_effect_mass}),  
and by the qualitative success of superfluid models to accommodate
observed features of glitches and their relaxation (e.g. see
\cite{link96:_therm_glitch} and references therein). 
For both the study of oscillation modes of superfluid neutron stars,
and the modelling of the particular glitch instability, it is
important to know the (quasi-) stationary initial state of the
unperturbed neutron star. The principal difference with ordinary-fluid
models consists of the fact that in a superfluid model there are
\emph{two fluids} (i.e. the superfluid neutrons and all the rest),
that can flow independently. As a result of the slow spindown process
of the neutron star, the neutron superfluid will lag behind the
charged constituents, due to its complete absence of viscosity. The
generic configuration therefore corresponds to a (quasi-) stationary
two-fluid model, with the two (strongly coupled!) fluids rotating
around the same axis, but with different rotation rates.  This model
has been studied before in both the Newtonian 
(\cite{prix99:_slowl_rotat_two_fl_ns,prix02:_slow_rot_ns_entrain}) 
and the relativistic framework
\cite{andersson01:_slowl_rotat_GR_superfl_NS}, using the
slow-rotating approximation and a simple  ``polytropic'' two-fluid
equation of state.  Here we describe a fully relativistic numerical
approach to solve the problem for arbitrary rotation rates.

\section{The two-fluid model}

We briefly present the notation and formalism used to describe
the two-fluid model for superfluid neutron stars, a more 
detailed discussion can be found in
\cite{carter98:_relat_supercond_superfl,langlois98:_differ_rotat_superfl_ns}.
We consider two fluids, representing the
neutrons (denoted by the label $\n$) and the remaining charged
components (denoted by $\p$) respectively. Superfluidity allows the 
neutrons to flow independently of the remaining constituents, 
without any direct frictional force between the neutrons and the
second fluid. Another implication of superfluidity is that the
rotating neutron fluid will be threaded by a lattice of microscopic
quantized vortices, but their 
effect on the macroscopic scale is neglected here, so we treat
the neutrons simply as an inviscid fluid.
The kinematics of the two fluids is described by the respective
rest-frame particle number densities $n_\n$ and $n_\p$, together
with the four-velocities $\vec{u_\n}$ and $\vec{u_\p}$.
The corresponding particle four-currents are therefore
\mbox{$\vec{n_\n}= n_\n \vec{u_\n}$} and \mbox{$\vec{n_\p}= n_\p
  \vec{u_\p}$}, and we assume strict conservation of both fluids
(i.e. we exclude the possibility of ``transfusion''
\cite{langlois98:_differ_rotat_superfl_ns} via $\beta$-reactions
between protons and neutrons), which means
\begin{equation}
  \nabla\cdot \vec{n_\n} = 0\,,\qaq 
  \nabla\cdot \vec{n_\p} = 0\,.
\end{equation}
The dynamics is governed by the total energy density $\E(n_\n, n_\p,
\Dv^2)$, i.e. ``equation of state'', where $\Dv$ represents the
relative velocity between the two fluids, and is defined as 
\be
\Dv^2 \equiv 1 - \left( {n_\n  n_\p  \over \vec{n_\n}\cdot\vec{n_\p}}\right)^2\,,
\fe
corresponding to a gamma factor $\Gam$ of the relative motion, which is
\be
\Gam = - \vec{u_\n}\cdot\vec{u_\p}  = \left(1 - \Dv^2\right)^{-1/2}\,.
\fe
The energy scalar $\E$ determines the first law of thermodynamics, namely
\be
d\E = \mu^\n \, d n_\n + \mu^\p \, d n_\p + \entr \, d\Dv^2\,,
\fe
which defines the chemical potentials $\mu^\n$ and $\mu^\p$ of the two
fluids, and the ``entrainment function'' $\entr$, which measures the
dependence of the energy density on the relative velocity.  
The equations of motion for the two interacting fluids are derived
from a ``convective'' variational principle
\cite{carter89:_covar_theor_conduc,langlois98:_differ_rotat_superfl_ns} 
based on the hydrodynamic Lagrangian \mbox{$\Lagr = -\E$}. This defines 
the canonical momenta  $\vec{p^\n}$ and $\vec{p^\p}$ via the total
differential 
\mbox{$\d\Lagr = \vec{p^\n} \cdot \d\vec{n_\n} + \vec{p^\p}\cdot \d\vec{n_\p}$}.
These canonical momenta are found explicitly as
\bea
\vec{p^\n} &=& \mu^\n \vec{u_\n} + {2 \entr \over n_\n \Gam^3} 
\left(\vec{u_\p} - \Gam \vec{u_\n} \right)\,,\\
\vec{p^\p} &=& \mu^\p \vec{u_\p} + {2 \entr \over n_\p \Gam^3} 
\left(\vec{u_\n} - \Gam \vec{u_\p} \right)\,,
\fea
which shows that the effect of the entrainment $\entr$ is to 
make the momenta deviate from their respective four-velocity
directions. This effect has first been discussed (in different terms)
by Andreev and Bashkin in the context of mixtures of superfluid $^4$He
and $^3$He \cite{andreev75:_three_velocity_hydro}, and is also
equivalent to a description in terms of ``effective masses'' 
\cite{prix02:_slow_rot_ns_entrain}. 
In the absence of mutual friction and pinning forces between 
the two fluids, the equations of motion have a very simple form,
despite the fact that two fluids are coupled via the equation of
state, namely
\bea
\vec{u_\n}\cdot d \vec{p^\n} = 0\,, \qaq
\vec{u_\p}\cdot d \vec{p^\p} = 0\,,
\label{eq:EOM}
\fea
where $d$ denotes the exterior derivative, i.e. 
\mbox{$(d\vec{p})_{\a\b} = \nabla_\a p_\b - \nabla_\b p_\a$}.
The corresponding two-fluid energy-momentum tensor $\tens{T}$ is
obtained from the variational principle as
\be
\tens{T} = \vec{n_\n}\otimes \vec{p^\n} + \vec{n_\p}\otimes\vec{p^\p}
+ \Press \tens{g}\,,
\label{eq:T}
\fe
where $\tens{g}$ is the metric tensor, and $\Press$ is the generalized
pressure, defined as 
\be
\Press + \E = n_\n \mu^\n + n_\p \mu^\p\,.
\fe

\section{Symmetries and integrals of motion}

We consider spacetimes that are stationary, axisymmetric and
asymptotically flat. Stationarity and axisymmetry are associated with
two (commuting) Killing vector fields \mbox{$\vec{t}$} and 
\mbox{$\vec{\varphi}$} respectively. We choose adapted
polar-type coordinates $\{t,r,\theta,\varphi\}$, which are such that
\mbox{$\vec{t}=\partial/ \partial t$} and 
\mbox{$\vec{\varphi}=\partial / \partial \varphi$}.
We assume the two fluids are in purely circular motion around
the $z$-axis (i.e. we exclude convective meridional currents), so we
write 
\be
\vec{u}_\n = \gamma_\n \left(\vec{t} + \Omega_\n \vec{\varphi} \right)\,,\qaq
\vec{u}_\p = \gamma_\p \left(\vec{t} + \Omega_\p \vec{\varphi} \right)\,,
\fe
where $\Omega_\n$ and $\Omega_\p$ are the respective rotation rates of
the two fluids.
The most general metric (in quasi-isotropic maximal slicing gauge) 
\cite{bonazzola93:_axisy} can be written as  
\be
ds^2 = -(N^2 - N_\varphi N^\varphi) \, dt^2 
- 2 N_\varphi \,d\varphi \,dt 
+ A^2\left( dr^2 + r^2 d\theta^2 \right) + B^2 r^2
\sin^2\!\theta \,d\varphi^2\,,
\fe
where $N$, $N^\varphi$, $A$ and $B$ are functions of $(r, \theta)$,
and \mbox{$N_\varphi \equiv g_{\varphi\varphi} N^\varphi$}.
With the additional assumption of \emph{uniform} rotations ($\nabla
\Omega=0$), the equations of motion (\ref{eq:EOM}) simplify greatly 
and can be shown to
reduce to two first integrals of motion 
\cite{andersson01:_slowl_rotat_GR_superfl_NS}, namely
\be
\mu^\n = \gamma_\n \, \mathrm{const}_\n \,,\qaq
\mu^\p = \gamma_\p \, \mathrm{const}_\p\,.
\label{eq:1stInt}
\fe

\section{Results and discussion}

Under the assumptions of the previous section, the Einstein equations
reduce to four coupled elliptic equations for the four unknown
functions $N,\,N^\varphi,\,A$ and $B$, with the source terms given 
by the energy-momentum tensor  (\ref{eq:T}). These can be solved very
accurately using an iteration scheme involving the first integrals
of motion (\ref{eq:1stInt}). The numerical scheme is based on a
multi-domain pseudo-spectral method, which is described in more detail
in \cite{bonazzola98:_numerical}, and which we have appropriately
extended to the two-fluid model considered here.
In principle the problem can be solved for any equation of state
(EOS), but currently only a polytropic toy-model EOS is implemented.
This, however, is sufficient to show the qualitative features of
a superfluid neutron star models as opposed to a single fluid
model. The EOS used here is a ``generalized polytrope'' of the form  
\be
\E = (m_\n n_\n + m_\p n_\p) c^2 + 
{1\over2} \kappa_\n n_\n^{\gamma_{1}} + 
{1\over2} \kappa_\p n_\p^{\gamma_{2}} + 
\kappa_{\n\p} n_\n^{\gamma_{3}} n_\p^{\gamma_{4}} + \beta 
n_\n^{\gamma_{5}} n_\p^{\gamma_{6}} \Dv^2\,.
\label{eq:EOS}
\fe
We note that the two fluids are coupled via a ``symmetry energy''-type
term proportional to $\kappa_{\n\p}$ and an ``entrainment'' term
proportional to $\beta \Dv^2$.

\begin{figure}[ht]
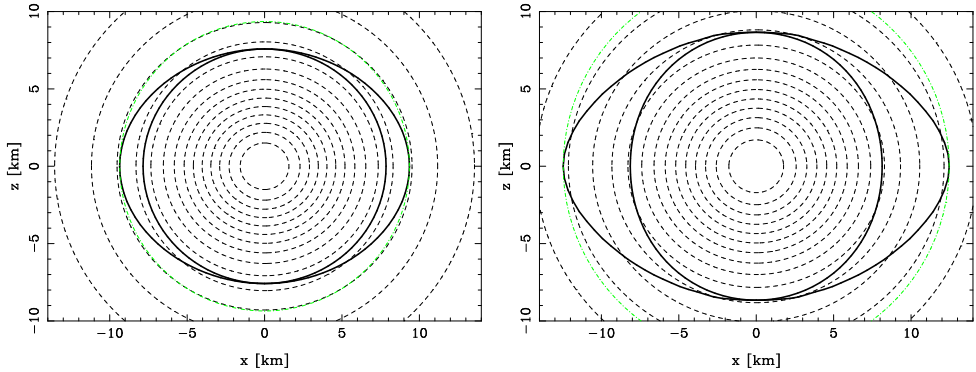

  \parbox{\textwidth}{
    {\includegraphics[angle=-90,width=0.49\textwidth]{fig1_ere}}\hspace{0.1cm}
    {\includegraphics[angle=-90,width=0.49\textwidth]{fig2_ere}}
  }
  \caption{Two-fluid configurations with $\Omega_\n =
    2000\,\pi~\mathrm{s}^{-1}$ and $\Omega_\p =
    200\,\pi~\mathrm{s}^{-1}$. The thick lines represent the surfaces of
    the two fluids respectively, while the dashed lines are the
    isosurfaces of ``gravitational potential'' $\ln N$.
    In the left-hand configuration, the two fluids are locally
    non-interacting, i.e. $\kappa_{\n\p}=0$ and $\beta=0$, while in the
    right-hand configuration we have chosen $\kappa_{\n\p}=0.02$ and $\beta = 0.12$.
  }
  \label{fig1}
\end{figure}

The numerical two-fluid code is currently in the test phase, internal
consistency checks like the relativistic viriel
(cf. \cite{bonazzola93:_axisy}) are satisfied to about $10^{-5}$ at a
convergence of $10^{-6}$. 
However, further tests are necessary, in particular comparisons with
previous results in the slow-rotation approximation, both relativistic
\cite{andersson01:_slowl_rotat_GR_superfl_NS} and Newtonian
\cite{prix99:_slowl_rotat_two_fl_ns,prix02:_slow_rot_ns_entrain}. 
The latter, in particular, allowed an analytic solution for a subclass of
the EOS (\ref{eq:EOS}). Furthermore, the existence of
\emph{prolate} solutions for the slower rotating fluid was found
analytically in \cite{prix02:_slow_rot_ns_entrain}, and we have
been able to reproduce these solutions with our numerical code (see
Fig.~\ref{fig1}). 
In Figure~\ref{fig1} we see the influence of 
the interaction (characterized by $\kappa_{\n\p}$ and $\beta$) between the
two fluids on the stationary configuration of a neutron stars model
with roughly $M\approx1.4\,M_\odot$ and $R\approx 11~\mathrm{km}$.
The EOS parameters used are $\gamma_{1}=\gamma_{2}=2$, and 
$\gamma_{3}=\gamma_{4}=\gamma_{5}=\gamma_{6}=1$, and $\kappa_\n =
\kappa_\p = 0.06$. 
In this configuration, one fluid is rotating at 
$\Omega_\n = 2000\,\pi~\mathrm{s}^{-1}$, while the second fluid
rotates much slower at only $\Omega_\p = 200\,\pi~\mathrm{s}^{-1}$. 
The left-hand graph shows the solution for two fluids which are
only gravitationally coupled (i.e. $\kappa_{\n\p}=\beta=0$), while in
the right-hand graph there is additional coupling via the
``symmetry''-term $\kappa_{\n\p}$ and entrainment $\beta$. The effect
is to  push the faster fluid further out and to ``squeeze'' the slower
fluid, in this case to the point of actually making it prolate. Both
the uncoupled and the coupled configuration are characterized by
the same central chemical potentials 
$\mu^\n(0) = \mu^\p(0) = 0.3 \,m c^2$, which is why their resulting
total masses and radii differ by about $10\,\%$. 

Apart from further quantitative tests, more work is necessary in order
to allow for more realistic equations of state, and to include the
possibility of vortex-mediated pinning forces between the two
fluids. 

\bigskip
{\textbf{Acknowledgements.} RP acknowledges support from the EU
Programme 'Improving the Human Research Potential and the
Socio-Economic Knowledge Base' (Research Training Network Contract
HPRN-CT-2000-00137).  GLC acknowledges partial support from NSF grant 
PHYS-0140138.}

\end{document}